\documentclass[aps,prb,reprint,superscriptaddress]{revtex4-1}

\usepackage{graphicx}
\usepackage{epstopdf}
\usepackage{dcolumn}
\usepackage[colorlinks]{hyperref}
\usepackage{verbatim}
\usepackage{amsmath}

\begin{document}

\title{Absence of superconductivity in NbB}

\author{F. Abud}\email{fabioabd@if.usp.br}
\affiliation{Instituto de F\'{i}sica, Universidade de S\~{a}o Paulo, Rua do Mat\~{a}o, 1371, 05508-090, S\~{a}o Paulo, SP, Brazil}

\author{L. E. Correa}
\affiliation{Escola de Engenharia de Lorena, Universidade de S\~{a}o Paulo, Caixa Postal 116, 12600-970, Lorena, SP, Brazil}

\author{I. R. Souza Filho}
\affiliation{Escola de Engenharia de Lorena, Universidade de S\~{a}o Paulo, Caixa Postal 116, 12600-970, Lorena, SP, Brazil}

\author{A. J. S. Machado}
\affiliation{Escola de Engenharia de Lorena, Universidade de S\~{a}o Paulo, Caixa Postal 116, 12600-970, Lorena, SP, Brazil}

\author{M. S. Torikachvili}
\affiliation{Department of Physics, San Diego State University, San Diego, California 92182-1233, USA}

\author{R. F. Jardim}
\affiliation{Instituto de F\'{i}sica, Universidade de S\~{a}o Paulo, Rua do Mat\~{a}o, 1371, 05508-090, S\~{a}o Paulo, SP, Brazil}

\date{\today}

\begin{abstract}

A systematic study of the superconducting properties in a series of arc-melted Nb-B samples close to the 1:1 composition was carried out. Powder X-ray diffraction (XRD) shows that all samples are both non-stoichiometric, and comprising of two crystal phases: a majority orthorhombic NbB-type phase, and traces of a minor body-centered cubic Nb-rich phase Nb$_{ss}$ with stoichiometry close to Nb$_{0.98}$B$_{0.02}$. The emergence of superconductivity near T$_c \sim$ 9.0 K was inferred from magnetization data in chunk and powder samples. However, the very small superconducting volume fractions are inconsistent with superconductivity arising from the major NbB phase. On the other hand, micrographs of selected samples clearly show that the minority Nb$_{ss}$ forms a three-dimensional network of filaments that meander around the grains of the majority phase, forming a percolation path. Here we report the superconductivity of the Nb$_{ss}$ phase, and argue that the low superconducting volume 
 fraction of non-stoichiometric NbB and zero resistance are due to the filaments of the minority phase. The electronic contribution to the entropy of the superconducting state, yielded from an analysis using the alpha model for single-band systems, indicates that the Sommerfeld constant of the arc-melted samples is close to the values found in non-superconducting NbB. Micrograph, XRD, and bulk measurements of magnetization, electrical resistivity, and specific heat suggest that the superconducting state in the NbB samples bearing some Nb$_{ss}$ minority phase is due to the latter.

\end{abstract}

\maketitle

\section{Introduction}

Superconductivity in borides has been widely studied for over six decades.\cite{Hulm1951_PR,Matthias1952_PR,Hardy1954_PR}
The more recent discovery of superconductivity in MgB$_2$,\cite{Nagamatsu2001_Nature} with superconducting critical temperature T$_c \sim$ 40 K renewed the interest in boron compounds, and provided the impetus behind the review of the properties of a large number of transition-metal boride.\cite{Lortz2006_PRB,Kawano2003_JPhysSocJpn,Kayhan2012_SolStaSci,Gou2013_PRL}
Frequently these studies indicated that the phase purity, non-stoichiometric compositions, and the presence of interstitial defects are important parameters in determining the physical properties in these materials.\cite{Buzea2001_SupScTech}
In this framework, the emergence of superconductivity in several borides such as TaB$_2$, NbB$_2$, and ZrB$_2$ has generated some controversy.\cite{Cooper1970_ProcNatSocUSA,Leyarovska1979_JoAC,Gasparov2001_JETP,Rosner2001_PRB}
Furthermore, a reported T$_c \sim$ 8.25 K in NbB,\cite{Akopov2017_AdvMat} was the highest amongst (TM)B compounds (TM = transition metal).
The potential for applications, and perhaps a starting point for the search for higher T$_c$ materials were identified as far back as 1952.\cite{Matthias1952_PR} Despite the large number of studies addressing the properties of the monoboride compounds, detailed studies of the superconducting properties of NbB are still lacking.

Previous studies of superconductivity in NbB may be regarded as controversial.
First, the value of T$_c$ has been reported in a broad range of temperature, ranging form 6 K\cite{Hulm1951_PR} to 8.25 K.\cite{Matthias1952_PR}
Secondly, other studies argued that superconductivity is absent down to 1.1 K \cite{Vandenberg1975_MatResBul} or even 0.42 K.\cite{Leyarovska1979_JoAC}
Besides these discrepancies, low-temperature specific heat data in NbB showed no evidence of superconductivity down to 1.5 K.\cite{Tyan1969_JPhysChemSol}
The lack of any features in the specific heat consistent with superconductivity down to 1.5 K, raises the question of whether the superconductivity identified by electrical transport and magnetic susceptibility is a bulk phenomenon.\cite{Tyan1969_JPhysChemSol}

The NbB phase crystallizes in the orthorhombic symmetry, space group Cmcm (no. 63) Pearson symbol oC8, with lattice parameters \textit{a} = 3.298, \textit{b} = 8.724, and \textit{c} = 3.166 \r{A}.\cite{Andersson1950AcChemSc}
The boron sites are arranged in chains along the $\it c$-axis, and located at the center of the faces of octahedra formed by Nb atoms.
In addition to these structural features, little attention had been paid to more specific properties of the superconducting state in NbB, and a detailed picture is lacking.

A number of different methods have been used for the synthesis of NbB: (i) arc melting;\cite{Matthias1952_PR,Andersson1950AcChemSc} (ii) solid state reaction;\cite{Tyan1969_JPhysChemSol,Brewer1951_JACS,Matsudaira1989_JoAC} and (iii) combustion synthesis.\cite{Yeh2006_JoAC} Regardless of the method, niobium-rich solid solutions, Nb$_{ss}$, and elemental Nb are commonly identified as extra phases.\cite{Brewer1951_JACS,Yeh2006_JoAC,Borges2003_JPhEq}
These Nb-rich solid solutions are known superconducting materials.\cite{DeSorbo1963_PR}
The partial substitution of B for Nb, the B presence in interstitial sites, and disorder, may all contribute to a drop in T$_c$.\cite{DeSorbo1963_PR}
Consequently, samples of NbB usually contain extra phases, making it difficult to discern precisely whether superconductivity is inherent to the NbB phase or not.

Within this context, we carried out a systematic study of the superconducting properties of arc-melted NbB samples.
The study was conducted in a set of twenty NbB specimens with chemical compositions in the vicinity of 1:1.
The samples were then characterized by powder XRD, scanning electron microscopy (SEM), and measurements of magnetization, electrical resistivity, and heat capacity.
Very small amounts of Nb$_{ss}$ were detected both by XRD and SEM imaging. Together with the measurements of the physical properties, this study sheds some light on the controversial superconductivity in NbB.

\section{Experimental}

Twenty NbB polycrystalline samples with composition close to 1:1 were prepared by arc melting in a Ti-gettered UHP argon atmosphere.
The starting materials for the $\sim 1$ gram samples were high purity Nb (99.99 at.\%) foil, and flakes of natural boron (99.5 at.\%; $^{10}$B/$^{11}$B $\sim 20/80~\%$) in atomic ratios close to 1:1.
Two samples were synthesized using isotopically enriched $^{10}$B and $^{11}$B, with chemical purities of 99.75 and 99.5 at.\%, respectively, and isotopic purities $>$ 99 \%. 
In order to avoid losses due to typical mini-explosions of B, the B flakes were first wrapped in the Nb foil, and the wraps were heated up slowly until the materials reacted fully. The resulting buttons were turned over and remelted at least four times in order to promote homogeneity. All samples presented small weight losses of $\sim$ 1 wt.\% after the melting process. If all losses were due to B, this would translate into a B atomic deficiency of about 10 \nolinebreak \%. In the set of samples discussed here no boron excess was added to compensate for the mass loss during the melting, as performed in other studies.\cite{Leyarovska1979_JoAC,Brewer1951_JACS} In order to promote and test for homogeneity, selected samples were annealed for $\sim$ 35 days at 1100 $^\circ$ C in evacuated quartz ampoules.

In order to verify the phase composition of our samples and calculate the lattice parameters we took XRD powder scans using a Bruker D8-Discover diffractometer. The measurements were performed at room temperature using CuK$_\alpha$ radiation in the $15 \leq 2\theta \leq 90 ^\circ$ range with a 0.05$^\circ$ step size, and 2 s counting time. Rietveld refinements of the XRD data were performed with FULLPROF.\cite{Rodriguezcarvajal1993_PhysicaB} Qualitative microstructural analysis was conducted using a LEO 1430 VP SEM (Zeiss) operating in the backscattered electrons mode, with 15 kV electron beam.

Magnetic susceptibility measurements, $\chi$ = M/H, taken on parallelepiped-shaped and powder samples, were performed using a Quantum Design superconducting quantum interference device (SQUID) magnetometer. The values of T$_c$ and superconducting volume fraction (SVF) were taken from the measurements of $\chi$. For the measurements of T$_c$ and $\chi$ the samples were first cooled to 1.8 K in zero field, a field was applied, and the $\chi$ vs T data was recorded upon warming to $\sim$ 15 K (ZFC) and cooling down again to 1.8 K (FC).

The temperature and magnetic field dependence of the electrical resistivity $\rho$(T,H) were obtained in small rectangular pieces with approximate dimension of $2~\text{x}~5~\text{x}~0.5~\text{mm}^{3}$, using a 4-probe technique, in a Quantum Design Dynacool Physical Property Measurement System (PPMS), operating in the T and H ranges of 1.8 - 300 K and 0 - 9 T, respectively. Specific heat C$_p$(T,H) data in the 0.05 to 300 K range were collected with the PPMS as well, with the measurements below 2 K being taken with a mating dilution refrigerator. A summary of the nominal compositions, boron isotope, heat treatment, identified phases, and T$_{c}$ values is shown in Table \ref{tab:table1}.

\begin{table*}
\caption{\label{tab:table1}Summary of the nominal compositions, heat treatments, identified phases, and T$_c$ values for arc-melted Nb-B samples. AC denotes as-cast samples and HT heat treated samples at 1100 $^\circ$C for 35 days in vacuum.}
\begin{ruledtabular}
\begin{tabular} { c c c c }
  Nominal Composition& Sample & Phases\footnotemark[1]& T$_c$\footnotemark[2] (K) \\
\hline
NbB & NbB \#1 - AC &NbB + Nb$_{ss}$ & 9.1\\ 

NbB & NbB \#1 - HT &NbB + Nb$_{ss}$ & 8.6 \\ 

NbB & NbB \#2 - AC &NbB + Nb$_{ss}$ & 8.9 \\ 

NbB & NbB \#2 - HT &NbB + Nb$_{ss}$ & 8.8 \\ 

NbB & NbB \#3 - AC &NbB + Nb$_{ss}$ & 9.1 \\ 

NbB & NbB \#4 - AC &NbB + Nb$_{ss}$ & 9.1 \\ 

NbB & NbB \#5 - AC &NbB + Nb$_{ss}$ & 8.9 \\ 

NbB & NbB \#6 - AC &NbB + Nb$_{ss}$ & 8.8 \\ 

NbB & NbB \#7 - HT &NbB + Nb$_{ss}$ & 8.5 \\ 

Nb$^{10}$B & NbB \#8 - AC &NbB + Nb$_{ss}$ & 8.9 \\ 

Nb$^{11}$B & NbB \#9 - AC &NbB + Nb$_{ss}$ & 8.9 \\ 

NbB$_{1.10}$ & NbB \#10 - AC &NbB + Nb$_5$B$_6$ + Nb$_{ss}$ & $\sim$ 6\footnotemark[3] \\ 

NbB$_{1.15}$ & NbB \#11 - AC &NbB + Nb$_{ss}$ & 9.0 \\ 

NbB$_{1.15}$ & NbB \#12 - AC &NbB + Nb$_{ss}$ & $\sim$ 6\footnotemark[3] \\ 

NbB$_{1.2}$ & NbB \#13 - AC &NbB + Nb$_{ss}$ & 9.0 \\ 

NbB$_{1.2}$ & NbB \#14 - AC &NbB + Nb$_{ss}$ & 9.0 \\ 

NbB$_{1.2}$ & NbB \#15 - AC &NbB + Nb$_5$B$_6$ + Nb$_{ss}$ & 8.8 \\ 

NbB$_{1.2}$ & NbB \#16 - AC &NbB + Nb$_5$B$_6$ + Nb$_{ss}$ & no sc for T $>$ 1.8 K \\ 

NbB$_{1.2}$ & NbB \#17 - AC &NbB + Nb$_5$B$_6$ + Nb$_{ss}$ & no sc for T $>$ 1.8 K \\ 

Nb$_{0.8}$B$_{0.2}$ & Nb-B-20 - AC & Nb$_{ss}$ + NbB  & 8.9 \\ 

Nb$_{0.9}$B$_{0.1}$ & Nb-B-10 - AC & Nb$_{ss}$ + NbB   & 8.9 \\ 

Nb$_{0.95}$B$_{0.05}$ & Nb-B-5 - AC & Nb$_{ss}$ + NbB   & 8.9 \\ 

\end{tabular}
\end{ruledtabular}
\footnotetext[1]{Identified on XRD scans.}
\footnotetext[2]{Defined as the onset of superconducting transition from the ZFC M(T) data under H = 5 Oe.}
\footnotetext[3]{Very broad and shallow superconducting transitions.}

\end{table*}

\section{Results and discussion}

\begin{figure}
\includegraphics[width=0.5\textwidth]{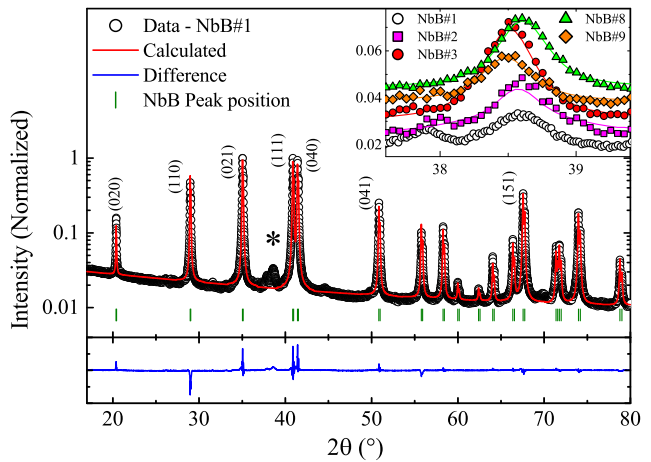}
\caption{\label{fig:RX}Typical normalized XRD $\theta$ - 2$\theta$ scans for as-cast NbB (NbB\#1). The red line represents a Rietveld refinement obtained with FULLPROF, and the difference plots is shown at the bottom. The reflection corresponding to the minority phase Nb$_{ss}$ solid solution is indicated with an asterisk. The inset displays an expanded view of XRD scans for five different samples near 2$\theta \sim$ 38.5 $^\circ$, the most intense reflection for Nb$_{ss}$.}
\end{figure}

A typical example of powder XRD scans for nearly all samples of this study is displayed in Fig. \ref{fig:RX}. The data shown is from an as-cast NbB sample (NbB\#1), in which we had the lowest mass loss upon melting ($\sim$ 0.5 wt. \%). A careful inspection of the diagram indicates that the most prominent diffraction peaks can be indexed using an orthorhombic structure with space group Cmcm (no. 63,  Pearson symbol oC8).\cite{Andersson1950AcChemSc} The calculated XRD matches closely the experimental pattern, as well as the scans found in the literature. The calculated lattice parameters yielded by the Rietveld refinement $\it a$ = 3.296, $\it b$ = 8.722, and $\it c$ = 3.165 \r{A}, are in excellent agreement with the literature.\cite{Andersson1950AcChemSc} Similar lattice parameters were found in all NbB samples studied in this work.

Although the mass loss measured in all NbB samples was rather small, it can be correlated to the appearance of a low-intensity Bragg reflection near 2$\theta \sim$ 38.5 $^\circ$ (indicated with an asterisk in Fig. \ref{fig:RX}). This impurity phase reflection occurs in a 2$\theta$ value very close to the (110) reflection of elemental Nb, suggesting the presence of a Nb-rich phase, possibly a solid solution. The correlation between the mass loss and the appearance of the (110) reflection indicate the presence of a Nb-rich Nb$_{ss}$ phase coexisting with the majority NbB phase, consistently with the Nb-B phase diagram.\cite{Okamoto2010_JPhaEqDif} 

In order to avoid the Nb-rich side of the Nb-B phase diagram, we studied a few samples containing a slightly overstoichiometric amount of B. However, the boron-rich Nb$_5$B$_6$ phase stabilized easily (see Table \ref{tab:table1}), which created a difficulty for our study. In light of the solubility limit of B in Nb close to the melting point being $\sim$ 2\%,\cite{Tang2008_Intermetallics} it is tempting to ascribe the composition of the Nb$_{ss}$ extra phase to being close to Nb$_{0.98}$B$_{0.02}$. Since the intensity of the XRD reflections is directly proportional to the concentration of the component producing it, we argue that the volume fraction of the Nb$_{ss}$ phase is rather low . However, given the right conditions, it may be responsible for filamentary superconductivity. 

\begin{figure}
\includegraphics[width=0.5\textwidth]{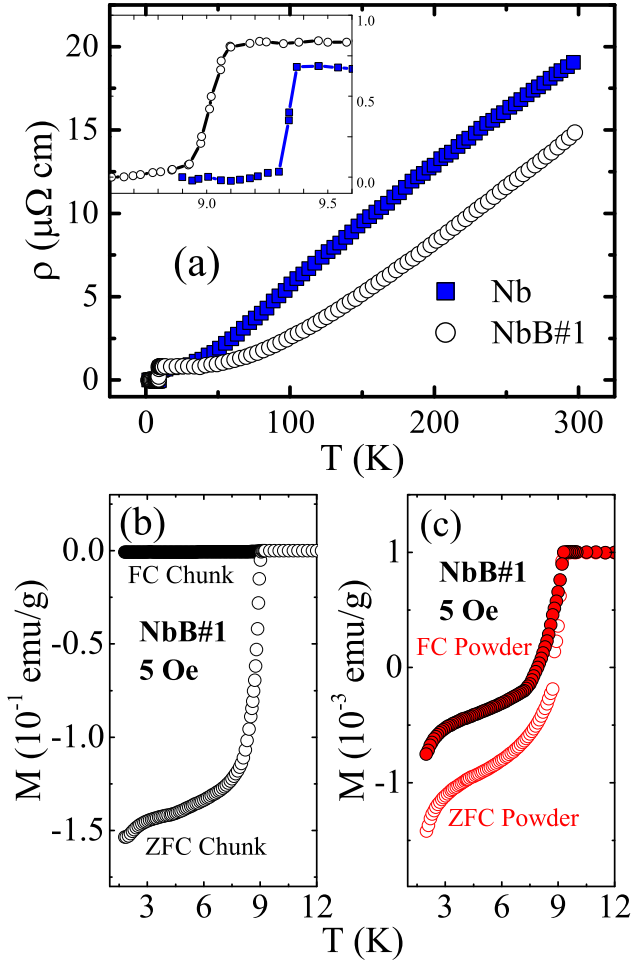}
\caption{\label{fig:R_M}(a) Temperature dependence of the electrical resistivity $\rho$(T) for NbB (NbB\#1), and the Nb foil material used as the starting material. The monotonic decrease of $\rho$(T) with temperature indicates metallic behavior for both samples. The superconducting transition region is exhibited in the inset of the figure. Field-cooled (FC) and zero-field cooled (ZFC) magnetization M(T) curves for an as-cast chunk (black symbols in (b)), and its powder (red symbols in (c)) of NbB (NbB\#1) under an external dc magnetic field of 5 Oe.}
\end{figure}

The temperature dependence of the electrical resistivity, $\rho$(T), of several NbB samples exhibited metallic behavior from ambient temperature down to $\sim$ 10 K, as displayed in the upper panel of Fig. \ref{fig:R_M}.

For a direct comparison of the $\rho$(T) data for the NbB sample and pure Nb, the $\rho$(T) data for the Nb foil starting materials is also displayed in Fig. \ref{fig:R_M}(a). The superconducting transitions for both NbB and pure Nb are clearly seen in the expanded view of the Fig. \ref{fig:R_M} inset. The value of T$_{c-\text{\tiny NbB}} = 9.0$ K for the NbB\#1 sample, estimated by using a 50 \% change in $\rho$(T) criterion, is very close to the value for nearly all other samples of this study (see Table \ref{tab:table1}). The width of the superconducting transition for the NbB\#1 sample is about $\sim$ 0.2 K, a value twice as large than that for the Nb foil. This difference does not rule out the possibility that the superconducting properties of the NbB samples could be due to elemental or Nb-rich solid solutions.

To gain some insight about the superconductivity in NbB, we carried our measurements of magnetization M(T) at 5 Oe, as shown in Fig. \ref{fig:R_M}(b). A weak paramagnetic signal is observed for temperatures above $\sim 10~\text{K}$, consistent with the metallic behavior of $\rho(T)$. The superconducting state is characterized by the diamagnetic signal observed in the ZFC data for $\text{T} < 9$ K. The measurements of M(T) in a small fragment of the as-cast NbB\#1 material yielded an onset for the superconducting transition near 9.1 K, in excellent agreement with resistivity data. A close inspection of the M vs T data indicates that the superconducting transition spans a relatively broad T-range below the onset; the value of M drops sharply from $\sim$ 9.1 to $\sim$ 8.1 K, and keeps dropping at a lower rate upon further cooling. However, no evidence for saturation down to 2 K for either chunks or powders were observed in any sample.

Neglecting the demagnetization factor of both phases, and taking the superconducting volume fraction (SVF) from the M vs T in 5 Oe, we estimated that values of SVF at 2 K to be lower than 0.5 \%, and $\sim$ 20 \% for the FC and ZFC measurements of all NbB samples, respectively. If the diamagnetic signal is originated in the NbB phase, the rather small SVF values are in conflict with the XRD data showing that NbB is the majority phase. Therefore the potential for superconductivity of any minority phase should be considered. The ZFC MxT data of Figs. \ref{fig:R_M}(b) and \ref{fig:R_M}(c) show that the diamagnetic signal for the large fragment of NbB is at least 2 orders of magnitude larger than for the powder, which gives credence to the conjecture that superconductivity could be originated in a minority phase. The percolation paths that support the screening currents of the ZFC data in the fragment lead to an overestimation of the SVF. Disruption of the percolation 
 by grinding leads to more realistic estimates of the SVF, which in turn is more consistent with minority phase superconductivity. We also acknowledge that a consideration of the XRD data with the superconductivity in the NbB samples leads to a conflict. First, the SVFs are rather low, and therefore inconsistent with arising from the majority phase as revealed in the XRD. However, $\rho$(T) drops to zero below T$_{c}$, suggesting that either the majority phase is superconducting, or that a minority phase is both superconducting and provide a percolation path.  

\begin{figure}
\includegraphics[width=0.49\textwidth]{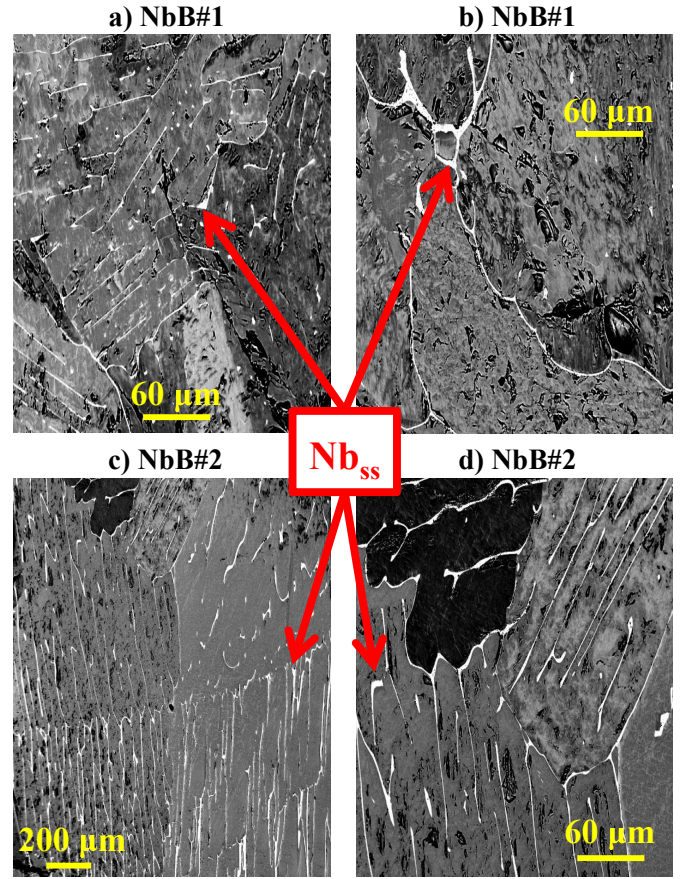}
\caption{\label{fig:Micro}Micrograph images for two NbB samples; upper and lower panels are for NbB\#1, and NbB\#2 samples, respectively. The red arrows point to white areas, which are identified with the Nb$_{ss}$ solid solution. The grayish areas are from the majority NbB phase.}
\end{figure}

In order to sort out whether the superconductivity in these polycrystalline NbB samples is in anyway related to the minority phases, we carried out detailed study of the microstructure. Displayed in Fig. \ref{fig:Micro} are surface micrographs of samples NbB\#1 and NbB\#2. The images clearly show the presence of two phases: NbB and Nb$_{ss}$, as indicated in Table \ref{tab:table1}. Since the image was built from the backscattered electrons, the lighter regions suggest a higher mean atomic number, and are associated with the Nb$_{ss}$ minority phase. The darker regions are associated with the NbB 1:1 majority phase. Filaments of the Nb$_{ss}$ minority phase coalesce on the grain boundaries, forming an interconnected network, as observed elsewhere.\cite{Borges2003_JPhEq} These filaments enclose large grains of the NbB matrix, forming loops or clusters with typical dimensions exceeding 10 \nolinebreak $\mu$m. The micromorphology of the arc-melted NbB samples can now be used 
 to address the problem of the low SVF.

Let's assume first that the superconductivity in the NbB samples arises from the Nb$_{ss}$ minority phase, with a morphology similar to a three-dimensional network of superconducting filaments. This assumption is consistent with several experimental findings in the arc-melted NbB samples:(i) the wide superconducting transition; (ii) the zero $\rho$(T) below T$_{c}$ due percolation through the filaments and linkage through the proximity effect;\cite{Kasumov1996_PRL, Likharev1979_RevModPhys} (iii) the very low SVF; and (iv) the suppression of the diamagnetic signal below T$_{c}$ by 2 orders of magnitude upon powdering the samples, presumably disrupting the filament network.\cite{[{See, for instance, }] Sathish2013_JPhysSocJpn} Upon fast cooling from the melt the Nb-rich solid solution phase is rapidly segregated to the edge of the grains where the filaments are formed. Therefore the filaments are quite susceptible to disorder, composition range, thinning, bottlenecking, and disruptions in continuity, leading to a variance in T$_{c}$ and to a broad superconducting transition. As indicated in the micrographs of Fig. \ref{fig:Micro}, the filaments have a large variance in thickness and display many interruptions; therefore proximity effect superconducting tunneling cannot be ruled out. 

\begin{figure}
\includegraphics[width=0.5\textwidth]{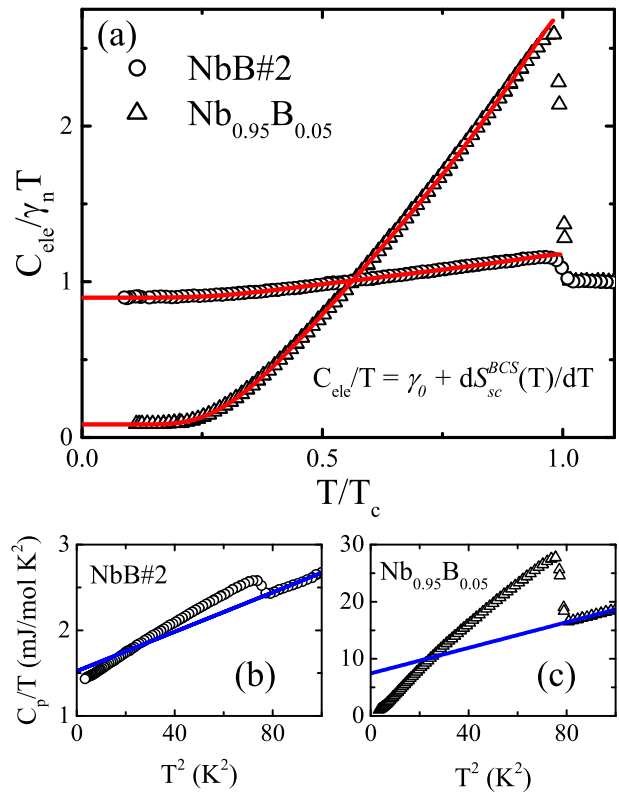}
\caption{\label{fig:Cp}(a) Reduced temperature T/T$_{c}$ dependence of the C$_{ele}$/$\gamma_n$T$_{c}$ ratio. The fit to the single-band model described in the text is represented by the red line in the main panel.(b) and (c) display C$_p$/T vs T$^2$ for samples NbB\#2 and Nb$_{0.95}$B$_{0.05}$, respectively. The blue lines are extrapolations from the normal state data, and their values at T$^2$ = 0 were taken as the Sommerfeld coefficients $\gamma_n$.}
\end{figure}

Specific heat C$_\text{p}$(T) measurements near T$_{c}$ can provide a quantitative mean for testing the validity of our SFV argument. The electronic contribution to the specific heat C$_\text{ele}/ \gamma T$ for two samples (NbB\#2 and Nb$_{0.95}$B$_{0.05}$) is displayed in Fig. \ref{fig:Cp}. The first point addressed here is the emergence of a jump at T/T$_c$ = 1 for both samples, consistent with the onset of superconductivity. The discontinuity at T/T$_c$ = 1 is much less pronounced in the NbB\#2 sample, suggesting that its SVF should be much smaller than for Nb$_{0.95}$B$_{0.05}$, a composition close to what we expect for the intergranular filaments.

To further evaluate the quantitative importance of the C$_\text{p}$(T) data, we'll start by assuming that the specific heat of the lattice C$_{latt}$ is well described by the Debye model C$_\text{latt}$ = $\beta_{3}$T$^3$ + $\beta_{5}$T$^5$ at low temperatures. The electronic contribution to C$_\text{p}$(T), C$_\text{ele}$ = $\gamma_n$T, is approximated by the free-electron model and therefore C$_\text{p}$ = C$_\text{ele}$ + C$_\text{latt}$ is used to describe the normal state data.\cite{Lortz2006_PRB} In order to suppress the superconducting transition from the background data, measurements of C$_\text{p}$(T) in the NbB\#2 sample were carried out both in H = 0 and 1 T. The latter is sufficient to lower T$_{c}$ below 4 K. The extrapolation of the C$_\text{p}$(T) data from the normal phase to low temperatures yields a Sommerfeld constant $\gamma_n = 1.52~\text{mJ/mol~K\ensuremath{^2}}$, as shown (blue line) in Fig. \ref{fig:Cp}(b). This value is close to the $\gamma_
 \text{\tiny NbB} = 1.39~\text{mJ/mol~K\ensuremath{^2}}$ reported for a non-superconducting NbB in Ref. \onlinecite{Tyan1969_JPhysChemSol}. On the other hand, specific heat C$_\text{p}$(T) measurements for Nb$_{0.95}$B$_{0.05}$ yielded $\gamma_n = 7.44 ~ \text{mJ/mol~K\ensuremath{^2}}$, fairly close to the reported value of $7.80~\text{mJ/mol~K\ensuremath{^2}}$ for pure niobium,\cite{Leupold1964_PR} as shown in Fig. \ref{fig:Cp}(c).

To make a direct comparison with the predictions of the BCS theory, the values of C$_\text{ele}$/T ratio were normalized by the Sommerfeld constant $\gamma_n$, and plotted as a function of the reduced temperature T/T$_c$ in the Fig. \ref{fig:Cp}(a). The discontinuity in the specific heat at T$_{c}$ for the NbB\#2 sample $\Delta \text{C$_\text{ele}$}/ \gamma_n T_c = 0.18$, corresponds to only $\sim$ 12\% of the expected 1.43 value for the BCS theory, assuming weak-coupling limit. On the other hand, a value much closer to the BCS value of 1.43 was obtained for the Nb$_{0.95}$B$_{0.05}$ sample ($\Delta \text{C$_\text{ele}$}/ \gamma_n T_c \sim  1.73$), consistent with nearly full SVF. Neglecting an anisotropic gap scenario, the small value of $\Delta \text{C$_\text{ele}$}/ \gamma_n T_c = 0.18$ suggests that the SVF of the NbB\#2 specimen is rather small, consistently with the low SVF found from the magnetization measurements. We also mention that the discontinuity at T$
 _{c}$ in the specific heat of the NbB\#2 sample was the largest among all the NbB specimens studied.

The upper panel of Fig. \ref{fig:Cp} shows that C$_\text{ele}$/$\gamma_n$T for the NbB\#2 sample assumes a very large and nearly constant value ($\sim$ 0.9) in the limit towards T $\rightarrow$ 0. This feature is commonly found in non-superconducting transition metal elements and metallic alloys, but quite different from the exponential decrease predicted in isotropic BCS theory.\cite{Bardeen1957_BCS} The unexpected large values of C$_\text{ele}$/$\gamma_n$T for NbB suggests either a partial gapping of the Fermi surface, or perhaps that NbB is non-superconducting at all. Motivated by this puzzling result, we carried out a study of the behavior of the electronic entropy in the superconducting state, using a model of independent fermions (alpha model) as proposed elsewhere.\cite{Padamsee1973_JLowTempPhys} This model has been successfully applied in multi-band and noncentrosymmetric superconducting materials.\cite{Jo2014_APLMat,Bouquet2001_PRL,Cirillo2015_PRB} For a single-band case and
  in close analogy with the BCS theory, the entropy of the superconducting state $S_{sc}^{BCS}$ is related to its electronic contribution C$_\text{ele}$ by the following equation:\cite{Padamsee1973_JLowTempPhys}\linebreak

\begin{equation}
\label{eq:1}
\begin{split}
\frac{\text{C}_\text{ele}}{\text{T}} & = \gamma_0 + \frac{d S_{sc}^{BCS}}{dT}, \\
S_{sc}^{BCS} & = -\gamma_nT_c\frac{6\alpha}{\pi^{2}}\int_{0}^{\infty}f ln f + (1-f)ln(1-f) dy.
\end{split}
\end{equation}

\noindent where $\gamma_0$ is a residual electronic contribution, $f(E)$ is the Fermi-Dirac function with quasiparticle energies \linebreak $E = \Delta_0\sqrt{y^2+\delta(T)^2}$, $y$ is the single-particle energy in the normal state, and $\delta(T)$ is the temperature dependence of the superconducting gap. The temperature dependence of an isotropic s-wave BCS gap $\delta(T) =$  \nolinebreak 1.76 \nolinebreak tanh$\{1.82[1.018(T_c/T - 1)]^{0.51}\}$ was utilized here as an approximation for the numerical solution of the BCS gap equation in the weak-coupling limit.\cite{Carrington2003_PhysC} The parameter $\alpha = \Delta_0$/$k_BT_c$ is adjustable and closely related to the superconducting coupling strength, which may assume values higher than predicted by the BCS weak-coupling limit $\alpha = 1.76$. 
 
The theoretical fits based on the alpha model are represented by the solid (red) lines in Fig. \ref{fig:Cp}(a), and show excellent agreement with the experimental data. For the NbB\#2 sample, a value of $\alpha \sim 1.92$ was then obtained, corresponding to a superconducting gap energy $\Delta_0^{NbB} \sim $\nolinebreak 1.44 $meV$. This value of $\Delta{_0}$ is slightly lower but in good agreement with the $\sim$ 1.54 and $\sim$ 1.52 $meV$ found in our Nb and Nb$_{0.95}$B$_{0.05}$ samples, respectively. The value of the superconducting energy gap of Nb obtained from tunneling and heat capacity measurements,\cite{Bose2005_PRL,Macvicar1968_JApP} are in the 1.42 $ \leq \Delta_0^{Nb} \leq$ 1.60 meV range, supporting the notion that the superconductivity in some NbB samples is due to the Nb$_{ss}$ impurity phase.

The metallic component of the NbB\#2 sample is also of interest and it may be explored using the results of the fitting procedure discussed above. The residual component $\gamma_0$ (see Eq. \ref{eq:1}) is associated with the metallic phase and has been used as an adjustable parameter in the fitting. The $\gamma_0 \sim  1.38~\text{mJ/mol~K\ensuremath{^2}}$ value obtained from the fitting is close to the $\gamma_n$ extrapolated from normal state specific heat ($\gamma_n \sim 1.52~\text{mJ/mol~K\ensuremath{^2}}$), suggesting that $\sim$ 10\% of the NbB sample is out of the stoichiometry. In fact, when $\gamma_0$ obtained from the fitting procedure is assumed to be the Sommerfeld constant for the metallic NbB phase, a much better agreement with the previously published data is achieved; a $\gamma_\text{\tiny NbB}\sim 1.39~\text{mJ/mol~K\ensuremath{^2}}$ value was reported for non-superconducting NbB.\cite{Tyan1969_JPhysChemSol} For comparison purposes, applying the fitting procedure from
  the alpha model to Nb$_{0.95}$B$_{0.05}$ yielded a residual component $\gamma_0$ as small as $\sim  0.6~\text{mJ/mol~K\ensuremath{^2}}$, a value corresponding to only $\sim$ 8\% of its $\gamma_n$, but in line with a sample with nearly 100\% superconducting volume fraction.

\section{Conclusions}

A systematic study in twenty arc-melted samples of NbB with stoichiometry close to 1:1 revealed superconductivity with T$_c$ $\sim$ 9 K in all of them. The arc-melted samples had weight losses lower than $\sim$ 1 wt.\%, which perhaps was enough to allow the formation of at least one additional phase, a Nb-rich solid solution Nb$_{ss}$ probably with composition close to Nb$_{0.98}$B$_{0.02}$. The presence of this minority phase was verified by powder XRD and SEM micrographs, which revealed the formation of a percolative path. Our measurements of SVF, electrical resistivity, and specific heat lead to a solid understanding that the superconductivity is associated not with the majority NbB phase but rather with the filaments of the minority solid solution Nb$_{ss}$. In this context, the morphology of the Nb$_{ss}$ phase seen in the micrographs, clearly showing the percolation paths, is consistent with the zero electrical resistivity state, even if this phase can be barely detected in XRD
  scans. The absence of superconductivity in the NbB phase, as discussed here, is consistent with two previous studies,\cite{Leyarovska1979_JoAC, Tyan1969_JPhysChemSol} further indicating that the synthesis method plays a crucial role on whether the samples will bear some Nb$_{ss}$ filaments and superconduct, or be impurity free and not exhibit superconductivity. In summary, the findings in arc-melted NbB samples suggest great difficulty in avoiding the formation of a minority Nb-rich Nb$_{ss}$ solid solution using this technique. A close inspection of the superconducting properties, XRD, and micrographs, suggest that the source of superconductivity is the percolative network of the Nb-rich Nb$_{ss}$ solid solutions that accumulate around the grain boundaries. 

\section{Acknowledgments}

R.F.J. and A.J.S.M. gratefully acknowledge Zachary Fisk for his hospitality at University of California, Irvine, where part of this work was done, and for enlightening discussions. We also have benefited from fruitful discussions with G. C. Coelho and C. A. Nunes. This work was supported by Brazil's agencies FAPESP (Grants No. 2013/07296-2, 2014/12401-2, and 2014/19245-6) and CNPq (Grants No. 444712/2014-3 and 306006/2015-4).

\end{document}